\documentclass[cpp,a4paper,fleqn%
]{w-art}
\usepackage{times,cite,w-thm}
\theoremstyle{plain}

\theoremstyle{definition}

\newcommand{\abs}[1]{\left\vert#1\right\vert}

\newcommand{\eps}{\varepsilon}

\newcommand{\phdag}{{\phantom{\dag}}}

\newcommand{\conj}{{*}}
\newcommand{\phconj}{{\phantom{\conj}}}

\newcommand{\phsubone}{{\phantom{1}}}

\newcommand{\NitrogenGroundStateTermSymbol}{\ensuremath{^1\Sigma_g^+}}
\newcommand{\NitrogenGroundState}{\ensuremath{N_2(\NitrogenGroundStateTermSymbol)}}

\newcommand{\NitrogenDominantMetastableStateTermSymbol}{\ensuremath{^3\Sigma_u^+}}
\newcommand{\NitrogenDominantMetastableState}{\ensuremath{N_2(\NitrogenDominantMetastableStateTermSymbol)}}

\newcommand{\NitrogenNegativeIonResonanceTermSymbol}{\ensuremath{^2\Pi_g}}
\newcommand{\NitrogenNegativeIonResonance}{\ensuremath{N_2^-(\NitrogenNegativeIonResonanceTermSymbol)}}

\newcommand{\RCTCaptureME}{V_{\vec{k}}}

\newcommand{\hsdt}{\hspace{-0.6em}}
\newcommand{\hsdtexp}{\hspace{-0.15em}}

\newcommand{\AlTwoOThree}{Al$_{\mathrm{2}}$O$_{\mathrm{3}}$}
\newcommand{\MgO}{MgO}
\newcommand{\LiF}{LiF}
\newcommand{\SiOTwo}{SiO$_{\mathrm{2}}$}

\usepackage[]{graphicx}
\begin{document}
\DOIsuffix{theDOIsuffix}
\Volume{46}
\Month{01}
\Year{2007}
\pagespan{1}{}
\Receiveddate{XXXX}
\Reviseddate{XXXX}
\Accepteddate{XXXX}
\Dateposted{XXXX}



\title[Wall charge]{Wall charge and potential from a microscopic point of view}


\author[Bronold]{F. X. Bronold\footnote{Corresponding
     author: e-mail:{\sf bronold@physik.uni-greifswald.de}}
     }\address{Institut f\"ur Physik,
     Ernst-Moritz-Arndt-Universit\"at Greifswald, D-17487 Greifswald, Germany}
\author[Fehske]{H. Fehske}
\author[Heinisch]{R. L. Heinisch}
\author[Marbach]{J. Marbach}
\begin{abstract}
Macroscopic objects floating in an ionized gas (plasma walls) accumulate electrons more efficiently 
than ions because the influx of electrons outruns the influx of ions. The floating potential acquired 
by plasma walls is thus negative with respect to the plasma potential. Until now plasma walls 
are typically treated as perfect absorbers for electrons and ions, irrespective of the microphysics 
at the surface responsible for charge deposition and extraction. This crude description,
sufficient for present day technological plasmas, will run into problems in
solid-state based gas discharges where, with continuing miniaturization, the wall becomes an 
integral part of the plasma device and the charge transfer across it has to be modelled more 
precisely. The purpose of this paper is to review our work, where we questioned the perfect absorber
model and initiated a microscopic description of the charge transfer across plasma walls, put it into 
perspective, and indicate directions for future research.

\end{abstract}
\maketitle                   





\section{Introduction}

All low-temperature gas discharges are bound by macroscopic 
objects. In contrast to magnetically confined, high-temperature plasmas they strongly interact with 
solids which either operate as electrodes, providing the break-down voltage,
or simply as floating walls, preventing the constituents of the plasma to disappear. The most 
fundamental manifestation of the solid-plasma interaction is the plasma sheath adjacent to an
unbiased, floating wall. It is an intrinsic electron-depleted region which solely arises 
because the plasma is bound by a solid~\cite{Franklin76}.

The sheath is the macroscopic indication of a microscopic charge transfer across the
plasma wall. Electrons are deposited in and extracted from the wall until a negative 
wall potential results which repels electrons and attracts ions such that quasi-stationarity
of the potential can be maintained. The microscopic understanding of this process is rather
rudimentary. Usually, it is assumed that all electrons and ions hitting the wall annihilate
instantly which is the same as to say that the wall is a perfect absorber and that 
at the wall the electron and ion influx balance. Most of the 
modeling of gas discharges (see, for instance, Loffhagen and Sigeneger~\cite{LS09} for a recent 
review) uses this boundary condition and leaves thus no room for the description of the charge 
transfer across the plasma wall. At best the wall is characterized by an electron-ion  
recombination coefficient and secondary electron emission coefficients for 
various impacting species. 

Clearly, the perfect absorber model implicitly assumes that for the phenomena occurring in 
the discharge the time and spatial scales of the charge transfer across the plasma wall 
are irrelevant and hence there is no need to track them~\cite{Franklin76}. How electrons 
are trapped in or at the wall, what their binding energy and residence time is, how and from what 
kind of electronic states they are released, and how the sheath potential merges with the 
surface potential of the wall are beyond the scope of this crude modeling of the plasma wall. 
In various novel bounded plasmas~\cite{Meichsner12} it seems to be however rewarding to pay more 
attention to these questions. In dusty plasmas~\cite{BBB09,PAB08,FIK05,Bonitz12,Schubert12}, for instance, 
the total amount of charge soaked up by the dust particles affects of course the overall characteristic 
of the discharge~\cite{BKS06} and should thus be known as precisely as possible. 
Likewise it is by now also well known~\cite{GMB02,WYB05,BMG05,SLP07,Bogaczyk12} that the wall 
charge plays an active role in determining the spatio-temporal structure of dielectric barrier 
discharges~\cite{Kogelschatz03} and microplasmas~\cite{BSE06,Kushner05}. A detailed 
understanding of the charge transfer across the plasma-dielectric interface promises therefore 
an improved control of this type of discharges. Finally and most fascinating, in solid-state based 
microdischarges~\cite{TWH11,DOL10}, the (biased) plasma wall becomes with continuing miniaturization 
even an integral part of the discharge and thus needs to be explicitly modelled.

Inspired by Emeleus and Coulter~\cite{EC87} as well as Behnke and coworkers~\cite{BBD97,GMB02}
who attempted to describe the dynamics of the wall charge and its coupling to the bulk plasma with 
phenomenological rate equations, characterized by sticking coefficients, residence times, and 
recombination and emission coefficients, we initiated in the framework of the TRR24 an effort 
to describe the plasma wall beyond the perfect absorber
model~\cite{BFKD08,BDF09,BHMF11}. With an eye on grain charging in dusty plasmas and the wall 
charge in dielectric barrier discharges we calculated -- as a first step -- for various uncharged 
metallic~\cite{BDF09} and dielectric~\cite{HBF10a,HBF10b,HBF11} surfaces electron sticking 
coefficients and desorption times, determined the distribution of the wall charge across the 
interface between a plasma and a floating dielectric surface~\cite{HBF12}, and investigated  
how electrons are extracted from dielectric surfaces via de-exciting metastable 
molecules~\cite{MBF11,MBF12}. Below we discuss the status of our work, put it into perspective, 
and indicate where it should go in the future.


\section{Build-up of the wall charge: Deposition of electrons}
\label{BuildUp}

In contrast to the assumptions of the perfect absorber model, an electron impinging 
on a solid surface is either reflected, inelastically scattered, or temporarily 
deposited to the surface. Possible trapping states (or sites) and hence residence times 
and penetration depths depend on its energy, the inelastic coupling to the elementary excitations 
of the surface driving energy relaxation, and the work function (electron affinity) of the material.

The surface physics just described can be encoded in a Hamiltonian. For a planar 
surface, with a potential which varies only perpendicularly to the surface,
and using the eigenstates of this potential as a basis~\cite{BDF09},
\begin{align}
 H=\sum_q E_q c_q^\dagger c_q
   +\sum_s E_s d_s^\dagger d_s
   +\sum_{s,q,q'}\langle q|\hat{V}_s|q'\rangle c_q^\dagger c_{q'}~,
\label{Hphysi}
\end{align}
where the first term describes the motion of the impinging electron in the surface 
potential, the second term denotes the motion of the elementary excitations of the wall 
responsible for energy relaxation, and the last term is the coupling of these 
excitations to the electron; $q$ and $s$ label the respective states and excitations.

The basis in which the Hamiltonian is written down, the elementary excitations causing energy relaxation, 
and the coupling $\hat{V}_s$ depend on the surface. Below we show results for dielectric surfaces where 
the elementary excitations are acoustic phonons and the surface potential is a truncated
image potential. This model is applicable to a dielectric surface with negative electron affinity, that 
is, a dielectric where the bottom of the conduction band is above the potential just outside the surface,
\MgO\ and \LiF\ being two important examples. For such a surface image states are stable and can thus host 
the approaching electron. In order to investigate the dependence of the trapping scenario on material 
parameters, for instance, the Debye frequency and the dielectric function, and to show trends we applied 
the model however also to dielectric materials with positive electron affinity~\cite{HBF10a,HBF10b,HBF11}.

In analogy to physisorption of neutral particles~\cite{KG86} we take the time evolution of 
the occupancy of bound surface states labelled by $n$ as a measure for temporary trapping. It satisfies a rate 
equation~\cite{HBF10a,HBF10b,HBF11},
\begin{align}
\frac{\mathrm{d}n_n(t)}{\mathrm{d}t}=\sum_{n^\prime}\left[W_{nn^\prime} n_{n^\prime}(t)-W_{n^\prime n} n_n(t) \right] 
-\sum_k W_{kn}n_n(t) + \sum_k \tau_t W_{nk} j_k \text{ ,}
\label{RateEq}
\end{align}
where the transition probabilities \(W_{\cdots}\) have to be calculated from (\ref{Hphysi}), $j_k$ denotes 
the stationary flux corresponding to a single electron in the unbound surface state $k$, and $\tau_t=2L/v_z$ 
is the traveling time through the surface
region where energy relaxation occurs. For relaxation happening in external surface states, the length $L$ 
can be absorbed in the transition probabilities and drops out in the limit $L\rightarrow\infty$ which can
be meaningfully taken in this case. If however energy relaxation takes place inside the wall, as in dielectrics
with positive electron affinity, $L$ is the penetration depth and has to be obtained from experiments or 
theoretically calculated~\cite{BHMF11}.

\begin{figure}
\begin{minipage}{72mm}
\includegraphics[width=\linewidth]{Fig1.eps}
\caption{Inverse of the electron residence time as a function of the surface temperature $T_s$ for an electron
thermalized with a \LiF\ surface (left panel) and $k-$averaged prompt and kinetic sticking coefficients (right panel)
for an electron approaching a \LiF\ surface at $T_s=300 K$ with a kinetic energy which is Boltzmann
distributed over the unbound surface states with average energy $k_BT_e$. The perfect 
absorber model implies $s_e=1$ and $\tau_e^{-1}=0$.}
\label{fig:1}
\end{minipage}
\hfil
\begin{minipage}{65mm}
\includegraphics[width=\linewidth]{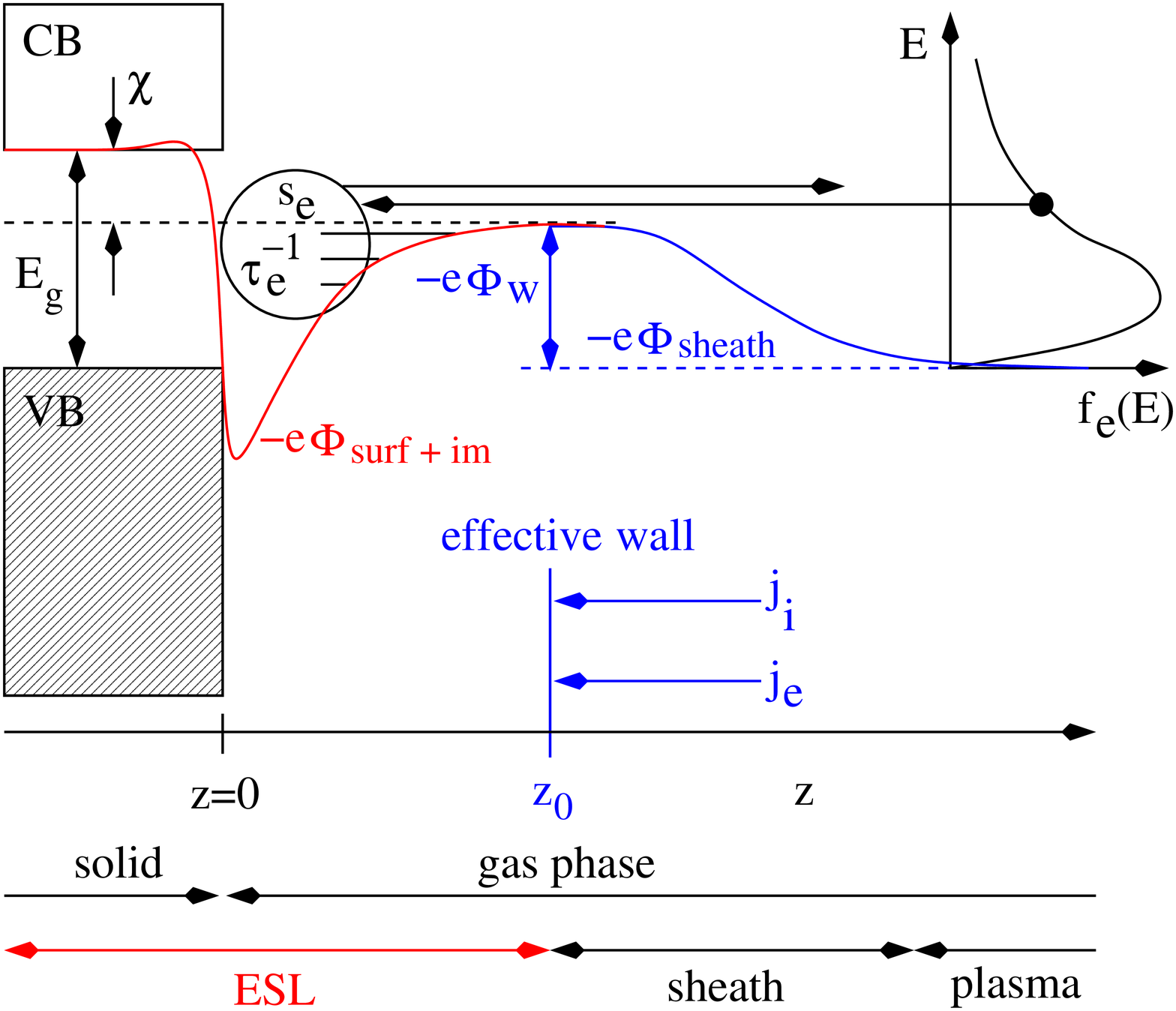}
\caption{Energetic situation in front of a dielectric surface
with $\chi<0$ and $-e\phi_w>0$. The red line is the
graded interface potential on which the model of an
electron surface layer (ESL) is based. Also shown is the electron
energy distribution function in the plasma, the effective wall where
the flux balance condition is enforced, and the
image states where the electrons comprising the wall charge get trapped.
}
\label{fig:2}
\end{minipage}
\end{figure}

A quantity characterizing trapping of an electron with wave number $k$ is the prompt 
sticking coefficient~\cite{KG86},
\begin{align}
s_{e,k}^\text{prompt}=\tau_t\sum_nW_{nk}~,
\end{align}
which is the probability to make during one round trip in the relevant surface region a transition 
from an unbound to a bound state. In case the surface potential supports only one bound state 
this is enough and the rate equation~(\ref{RateEq}) is not needed. If however the surface potential 
supports many bound states and the elementary excitations causing energy relaxation are not energetic 
enough to connect the lowest bound state directly with unbound states, 
the characterization of trapping has to be based on the rate equation. From its slowly 
varying part~\cite{KG86},
\begin{align}
\frac{\mathrm{d} n^\mathrm{slow}(t)}{\mathrm{d}t}=\sum_k s_{e,k}^\text{kin} j_k(t) -\frac{1}{\tau_e}n^\mathrm{slow}(t) \text{ ,} 
\end{align}
the residence or desorption time \(\tau_e\) and the kinetic sticking coefficient
\(s_{e,k}^\text{kin}\), which is the probability for trapping and relaxation of an electron with 
wave number k in the manifold of bound surface states, can then be extracted.

We applied the approach just outlined to various uncharged dielectric surfaces assuming electron 
physisorption to occur in the image potential, which is, as pointed out, rigorously true only for 
dielectrics with negative electron affinity~\cite{HBF10a,HBF10b,HBF11}. Electron energy relaxation 
at these surfaces is driven by acoustic phonons whose Debye energy is very often not only too small 
to connect the lowest bound state to the unbound states but also too small to connect the two lowest bound 
states. The physisorption kinetics of electrons at dielectric surfaces involves therefore multiphonon 
transitions~\cite{HBF10a,HBF10b,HBF11}. Only the initial trapping of electrons in the upper bound 
states, characterized by the prompt sticking coefficient, occurs via one-phonon processes. Multiphonon
processes contribute very little to it. Hence, initial trapping is rather insensitive to the surface
temperature. Relaxation after initial trapping depends on the strength of transitions from the upper bound 
states to the lowest bound state. If the lowest two bound states are linked by a one-phonon transition, a 
trapped electron relaxes for all surface temperatures, if a multiphonon process is required, the electron 
relaxes only for low temperatures whereas for room temperature and higher relaxation is inhibited leading
to a relaxation bottleneck. The 
dominant desorption channel depends also on the depth of the potential. For a shallow potential desorption 
occurs directly from the lowest bound state to the continuum. For deeper potentials desorption proceeds via 
the upper bound states. Desorption occurs then via a cascade in systems without and as a 
one-way process in systems with relaxation bottleneck~\cite{HBF11}.
 
Electron physisorption at an uncharged dielectric surface is thus an intriguing phenomenon. For the plasma
context most important is that \(s_e \ll 1\) and \(\tau_e^{-1}\neq 0\)~\cite{HBF10a,HBF10b,HBF11}, implying
that an initially uncharged dielectric surface with a negative electron affinity is not a perfect absorber 
for electrons. Representative results for \LiF\ are shown in Fig.~\ref{fig:1}. Since the wall charge does 
not affect the relative line up of the potential just outside the dielectric and the bottom of the 
conduction band (potential just inside the dielectric), as mistakenly assumed in~\cite{BHMF11} but corrected 
in~\cite{HBF12}, this conclusion also holds 
for a charged dielectric wall with negative electron affinity. The sticking coefficient is only larger and thus 
closer to the value implied by the perfect absorber model when the electron affinity is positive and the impinging 
electron enters the wall, that is, for materials with positive electron affinity. In such a case the residence 
time should be also much longer, as implied by the perfect absorber model. 

After the initial charge-up is completed, the wall carries a quasi-stationary negative charge, that is, 
in the notation of electron physisorption, a quasi-stationary electron adsorbate. We will now discuss 
how the spatial profile of the electron adsorbate normal to the crystallographic interface can 
be determined. The basic idea~\cite{HBF12} is to use a graded interface potential~\cite{Stern78} to 
interpolate between the sheath and the wall potential and to distribute the surplus electrons making 
up the wall charge in this potential under the assumption that at quasi-stationarity 
they are thermalized with the wall~\cite{TD85}. For the present purpose it suffices to describe 
qualitatively the simplest implementation of this idea -- the crude electron surface layer (see Fig.~\ref{fig:2} 
and~\cite{HBF12}). 

\begin{figure}[t]
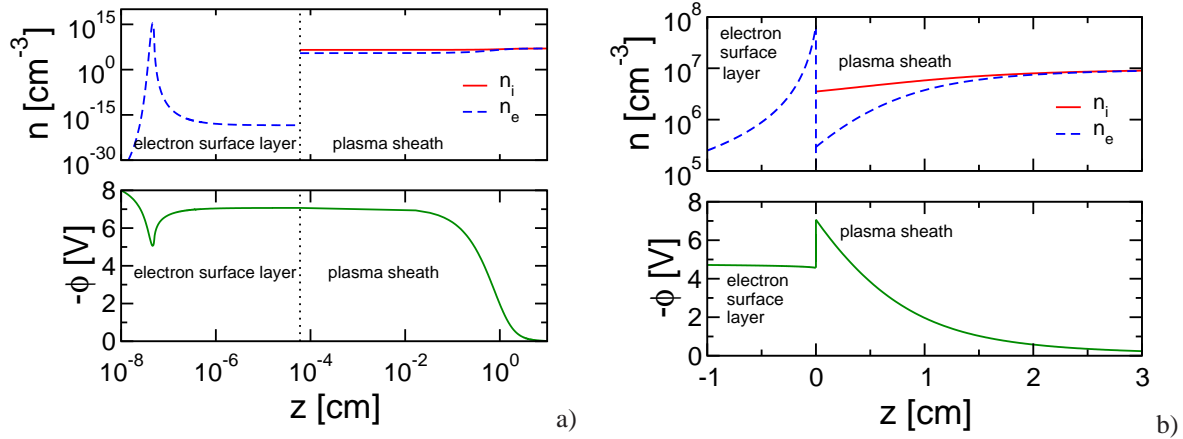

\begin{center}
\begin{minipage}{0.5\linewidth}
\includegraphics[width=0.9\linewidth]{Fig3a.eps}~a)
\end{minipage}\begin{minipage}{0.5\linewidth}
\includegraphics[width=0.9\linewidth]{Fig3b.eps}~b)
\end{minipage}
\caption{Density of plasma-supplied surplus electrons trapped in the ESL, electron
and ion density in the plasma sheath, and electric potential for a \LiF\ (panel \textbf{a}) and
an \AlTwoOThree\ surface (panel \textbf{b}) in contact with a helium discharge with plasma
parameters $n_0=10^7 cm^{-3}$ and $k_BT_e=2 eV$. The crystallographic interface is at $z=0$. 
Note, the different scales of the two
panels. The deep penetration of the \AlTwoOThree\ wall charge is due to the neglect of
defect states and other collision centers.}
\label{fig:3}
\end{center}
\end{figure}

The adsorbed electrons form an interface-specific electron distribution -- the electron surface layer 
-- across the planar interface at \(z=0\). Their spatial profile $n(z)$ can be calculated as 
follows~\cite{HBF12}. (i) First, an effective wall has to be defined. Its position, \(z_0>0\), marks 
the point where the sheath merges with the electron surface layer. For \(z>z_0\) electrons are repelled 
back into the plasma, whereas for \(z<z_0\) electrons are pushed towards the surface. Hence, a flux
balance is taken at \(z_0\). Moreover, the field 
strength at \(z_0\) due to the positive sheath charge is related to the total number of surplus 
electrons per unit area $N$. In order to calculate \(z_0\) and the field strength at \(z_0\) 
a sheath model and a flux balance condition are required. The results presented in Fig.~\ref{fig:3}
and discussed below, for instance, are -- for simplicity -- based on a collisionless sheath with a 
perfect absorber condition at \(z_0\). (ii) Second, equations for the electron distribution 
\(n(z)\) and the potential \(\phi(z)\) in the electron surface layer, that is, for \(z<z_0\) have
to be set-up. For that purpose, density functional theory can be employed~\cite{TD85}. The central 
equation is then given by minimizing the grand canonical potential of the interacting surplus electrons 
in the external potential provided by the surface and the sheath. In the local density approximation, it
reads~\cite{HBF12}
\begin{align}
-e(\phi_\mathrm{im}(z)+\phi_\mathrm{surf}(z)+\phi_\mathrm{C}(z))+\mu^h(n(z),T_s)-\mu=0~, 
\label{ESLcentral}
\end{align}
where \(\mu^h(n(z),T_s)\) is the chemical potential of a homogeneous electron gas with density 
\( n(z)\) at the surface temperature $T_s$, \(\phi_\mathrm{im}(z)\) is the graded image potential, 
\(\phi_\mathrm{surf}(z)\) is the graded surface potential accounting for the electron affinity of 
the surface, that is, the offset of the potential just outside the dielectric and the bottom of the 
conduction band (see Fig.~\ref{fig:2}), and \(\phi_\mathrm{C}(z)\) is the Coulomb potential satisfying the Poisson 
equation,
\begin{align}
\frac{\mathrm{d}}{\mathrm{d}z}\left(\epsilon(z)\frac{\mathrm{d}}{\mathrm{d}z}\phi_\mathrm{C}(z) \right)=4\pi e n(z)
\text{ ,}
\end{align}
with \(\epsilon(z)\) the graded dielectric constant interpolating between the dielectric constants of the
plasma and the dielectric. The total potential \(\phi(z)=\phi_\mathrm{im}(z)+\phi_\mathrm{surf}(z)+\phi_\mathrm{C}(z)\)
and \(\mu\) is the chemical potential of the surplus electrons.
(iii) Third, Eq.~(\ref{ESLcentral}) has to be solved iteratively (until \(\mu\) is stationary) subject to 
the boundary condition described in (i) and the additional constraint \(\int_{z_0}^{z_s} \mathrm{d}z \text{ }n(z)=N\) 
which guarantees charge neutrality between the electron surface layer and the plasma sheath. In the crude electron 
surface layer \(z_s<0\) is a cut-off which has to be chosen large enough in order not to affect the numerical 
results. In the refined electron surface layer it is the point where the electron surface layer merges with the 
intrinsic region of the dielectric~\cite{HBF12}.
 
In Fig.~\ref{fig:3} we show results for two floating dielectric walls in contact 
with a helium discharge with plasma density \(n_0=10^7 cm^{-3}\) and electron temperature \(k_BT_e=2eV\). 
The numerical calculations were performed as described in Ref.~\cite{HBF12}.
Depending on the electron affinity \(\chi\) the distribution of electrons at the surface assumes 
two distinct forms: For \LiF\ ($\chi<0$), the conduction band minimum lies above the potential just 
outside and the surplus electrons are bound in the image potential in front of the wall forming an 
external, very narrow surface charge which can be regarded as the quasi two-dimensional electron gas 
anticipated by Emeleus and Coulter~\cite{EC87}. Its spatial profile does not change much with surface 
temperature and surface density of electrons (depending on the plasma parameters). For \AlTwoOThree\
(\(\chi>0\)), the conduction band minimum lies below the potential just outside and the surplus 
electrons accumulate inside the dielectric forming an internal wall charge. Increasing the surface 
density of electrons (through the plasma parameters) makes the electron distribution more concentrated 
at the interface (steeple-like) while at higher surface temperature it is more extended. Also shown
in Fig.~\ref{fig:3} are the potential, the electron density, and the ion density in the sheath. The latter
two are discontinuous at \(z_0\) because inside the electron surface layer electron and ion fluxes 
are neglected. For further discussion see Ref.~\cite{HBF12}.

The model of an electron surface layer is an attempt to describe that part of the plasma boundary which 
leaks into the plasma wall. It provides a way to determine the distribution and binding energy of 
the wall charge as well as the spatial profile of the potential inside the wall. We described the 
electron surface layer for a floating wall but it can be generalized to a biased wall as well, that is, 
an electrode by simply supplementing it by an external bias.


\section{Tapping the charge of the wall: Extraction of electrons}

We now turn our attention to the extraction of electrons from the wall. The most important
process extracting electrons from the wall is the wall recombination of positive 
ions. But high energy electrons and metastable molecules or radicals carrying internal energy 
are also very efficient in releasing electrons from the wall. 

In the following we will focus on secondary electron emission from dielectric walls due to 
metastable nitrogen molecules. This process, which has been shown to stabilize the diffusive
mode of dielectric barrier discharges~\cite{BMG05}, plays an important role in an in-house
experimental effort to obtain, via surface and volume diagnostics, a complete characterization
of such discharges, comprising volume as well as surface processes~\cite{Bogaczyk12}.

\begin{figure}[t]
\begin{center}
\begin{minipage}{0.5\linewidth}
\includegraphics[width=0.9\linewidth]{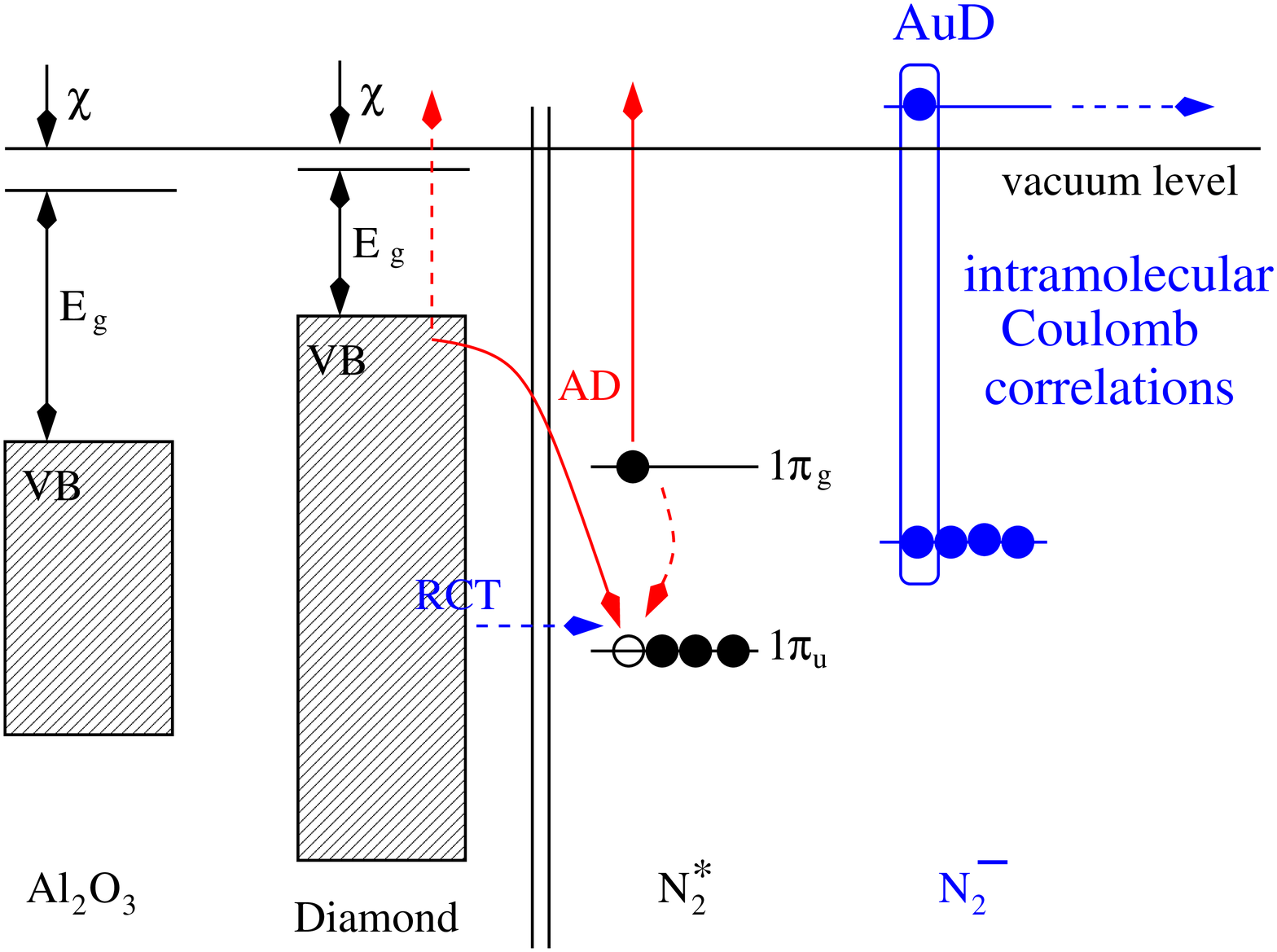}~a)
\end{minipage}\begin{minipage}{0.5\linewidth}
\includegraphics[width=0.9\linewidth]{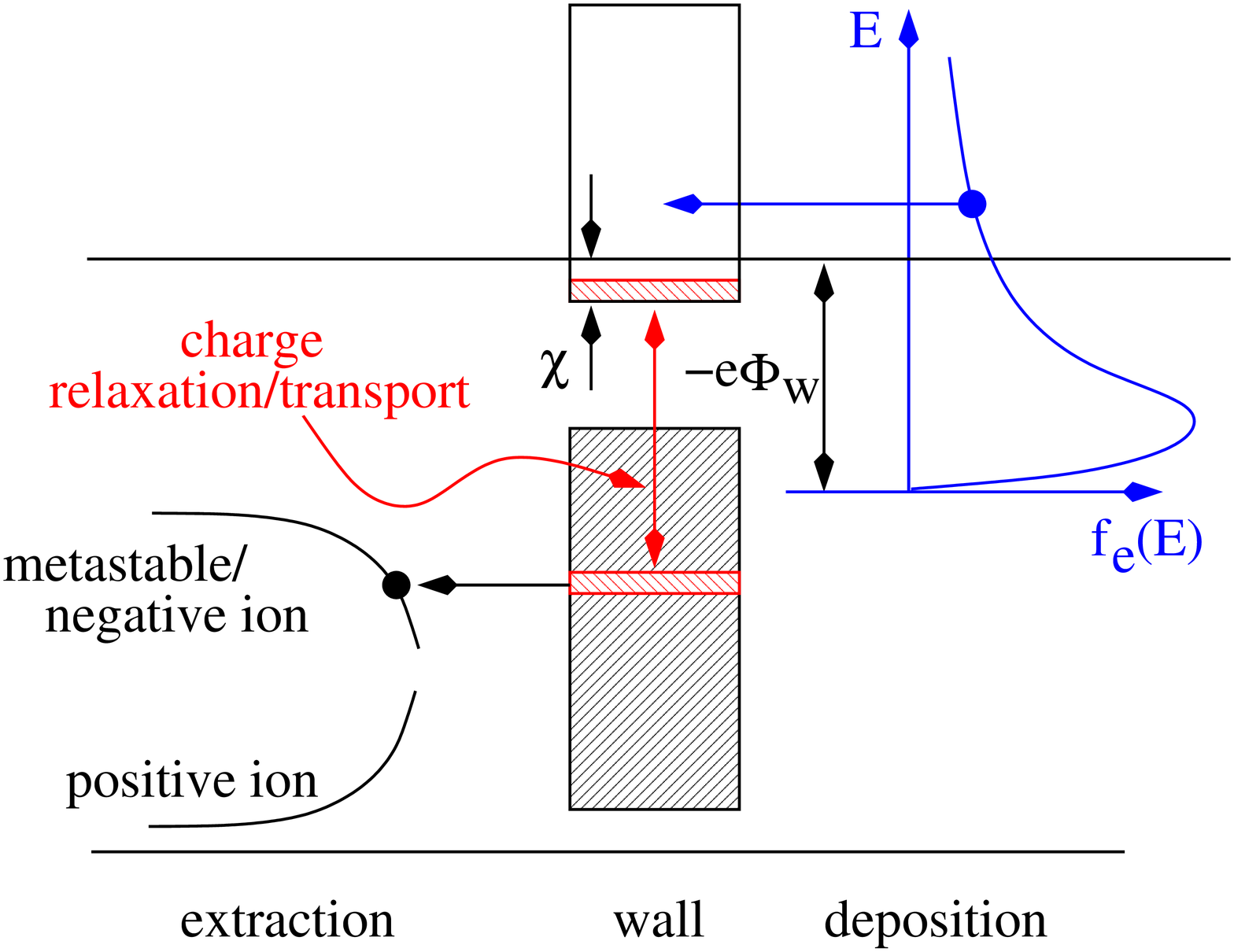}~b)
\end{minipage}
\caption{\textbf{a}) Energy scheme showing for a \NitrogenDominantMetastableState\ molecule 
scattering off a diamond and \AlTwoOThree\ surface direct (red dashed) and indirect (red solid)
Auger de-excitation (AD) and resonant charge transfer (RCT) with subsequent auto-detachment (AuD)
of the $\NitrogenNegativeIonResonance$  shape resonance (dashed blue).
\textbf{b}) Illustration of the energetic situation for electron deposition and extraction at 
a dielectric plasma wall with positive electron affinity. The electrons deposited in the 
conduction band of the solid are qualitatively shown as well as the holes in the valence band 
arising from the extraction of electrons by positive ions and/or metastables/negative ions. The 
inversion of the occupancy of the states in the dielectric may persist only temporarily
until it is eliminated by charge relaxation and transport or radiative processes.}
\label{fig:4}
\end{center}
\end{figure}

An important parameter for the modeling of dielectric barrier discharges is the secondary electron 
emission coefficient, characterizing the efficiency with which electrons can be extracted from 
the dielectric coverage of the electrode. It depends on the surface material, the projectile, and
the particular collision process. For \NitrogenDominantMetastableState\ two de-excitation 
channels~\cite{Stracke98} can provide an additional electron for the discharge. The molecule either 
de-excites via an Auger process corresponding to the reaction 
(possible at metallic surfaces~\cite{MBF11})
\begin{equation}
        \NitrogenDominantMetastableState + e_{\vec{k}} \rightarrow \NitrogenGroundState + e_{\vec{q}} \;,
        \label{eq-auger-reaction}
\end{equation}
with~$e_{\vec{k}}$ and~$e_{\vec{q}}$ denoting a surface electron and a free electron, 
respectively, or via a resonant electron capture leading to a negatively charged
shape resonance, $\NitrogenNegativeIonResonance$, which subsequently decays corresponding to
\begin{equation}
        \NitrogenDominantMetastableState + e_{\vec{k}} \rightarrow \NitrogenNegativeIonResonance
\rightarrow \NitrogenGroundState + e_{\vec{q}} \;.
        \label{eq-rct-reaction}
\end{equation}
For most dielectrics the Auger process is energetically suppressed, diamond being an important 
exception (see Fig.~\ref{fig:4}\textbf{a}). The combination of charge capture and subsequent 
decay of the negative ion (surface-induced as well as natural via auto-detachment) is thus the 
dominant de-excitation channel for \NitrogenDominantMetastableState\ at dielectric surfaces~\cite{MBF12}.

The reaction chain~(\ref{eq-rct-reaction}) consists of two sequential resonant tunneling processes (see 
Fig.~\ref{fig:4}\textbf{a}): resonant charge transfer (RCT) and auto-detachment (AuD). Each step can be modelled 
by an Anderson-Newns Hamiltonian~\cite{Yoshimori86},
which is an effective model, stripped of some of the microscopic details and characterized instead by a 
small number of material parameters which can be easily obtained. Models of this type are well suited for 
describing elementary processes at plasma walls where the lack of surface diagnostics prevents a more 
refined modeling. The Hamiltonian appropriate for~(\ref{eq-rct-reaction}) reads
\begin{equation}
H(t) = \sum_{\vec k} \eps_{\vec k}^{\phdag} \, c_{\vec{k}}^{\dag} \, c_{\vec{k}}^{\phdag} 
+ \eps_{m}^{\phdag}(t) \, c_{m}^{\dag} \, c_{m}^{\phdag}
+ \sum_{\vec k} \left( \RCTCaptureME^{\phconj\!}(t) \, c_{\vec{k}}^{\dag} \, c_{m}^{\phdag}
+ \RCTCaptureME^\conj(t) \, c_{m}^{\dag} \, c_{\vec{k}}^{\phdag} \right) \;,
\label{HamiltonianRCT}%
\end{equation}
where, for the first (second) step of the reaction chain~$\vec{k}$ denotes electronic states within the 
solid's valence band (free electron states) and~$m$ labels the lower (upper) ionization level of the
negative shape resonance~$\NitrogenNegativeIonResonance$. Using Keldysh Green functions~\cite{Blandin76}
it is possible to calculate from~(\ref{HamiltonianRCT}) the rate for resonant electron 
capture $\Gamma_0(t)$ and the rate for surface-induced decay $\Gamma_\mathrm{surf}(t)$. For details 
see Ref.~\cite{MBF12}. 

The ionization levels $\eps_m(t)$ and the tunnel matrix element~$\RCTCaptureME (t)$ depend on time via the
molecule's position relative to the surface. Assuming, for simplicity, normal incidence, and the molecule 
to start moving at $t_0=-\infty$ and to hit the surface at $t=0$, the trajectory of the molecule's center 
of mass is $\vec{R}(t) = z_R(t) \, \vec{e}_z = \left( v_0 \abs{t} + z_1 \right) \vec{e}_z\;,$
where $z_1$ is the turning point in the surface potential of the molecule and $v_0$ is the molecule's 
velocity. Explicit expressions for the matrix elements appearing in~(\ref{HamiltonianRCT}) can be found in 
Ref.~\cite{MBF12}. In Fig.~\ref{fig:4}\textbf{a} we show the relevant ionization levels of 
\NitrogenDominantMetastableState\ and $\NitrogenNegativeIonResonance$ for $t_0=-\infty$. Note, due to 
intra-molecular Coulomb correlations the ionization levels of $\NitrogenNegativeIonResonance$ are shifted 
with respect to the ones of \NitrogenDominantMetastableState\ . In accordance to the fact that 
$\NitrogenNegativeIonResonance$ is an unstable shape resonance, the upper ionization level of 
$\NitrogenNegativeIonResonance$ is above the vacuum level. The rate for the natural decay $\Gamma_\mathrm{nat}$ 
can be deduced from the life time assuming a Breit-Wigner-type line shape for the auto-ionization 
process~\cite{MBF12}. 

Relating the occupancies of the lower ($m=0$) and the upper ionization 
level ($m=1$) of $\NitrogenNegativeIonResonance$ to the fractions of metastable molecules $n_*(t)$,
negative ions $n_-(t)$, and ground state molecules $n_g(t)$, reaction~\eqref{eq-rct-reaction} can 
be coded into a system of ordinary differential equations whose solution gives~\cite{MBF12},
\begin{align}
  n_*(t) & = e^{-\int_{t_0}^{t_\phsubone} \hsdtexp\mathrm{d}t_1 \; \Gamma_0(t_1)} ~\;,~~ 
  n_-(t) = -\int_{t_0}^{t_\phsubone} \hsdt\mathrm{d}t_1 \; \frac{\mathrm{d} n_*(t_1)}{\mathrm{d}t_1} \; 
  e^{-\int_{t_1}^{t_\phsubone} \hsdtexp\mathrm{d}t_2 \; \Gamma_1(t_2)} \;,
\end{align}%
where $\Gamma_1=\Gamma_\mathrm{surf}+\Gamma_\mathrm{nat}$ is the total decay rate containing the surface-induced and the natural
decay. The fraction of ground state molecules follows from $n_g(t)=1-n_*(t)-n_-(t)$. Writing the total
decay rate $\Gamma_1$ in terms of an energy spectral function~\cite{MBF12} we also obtain the probability 
of emitting an electron,
%
\begin{equation}
  n(t) = \int_0^{\infty} \hsdt\mathrm{d}\eps_{\vec{q}}^\infty \int_{t_0}^{t_\phsubone} \hsdt\mathrm{d}t_1 \; 
  \varrho_1(\eps_{\vec{q}}^\infty,t_1) \, n_-(t_1) \;,
  \label{eq-consecutive-spectrum}
\end{equation}
which defines the secondary electron emission coefficient, $\gamma_e=n(\infty)$, as well as the 
energy spectrum of the emitted electron, $\mathrm{d}n(\infty)/\mathrm{d}\eps_{\vec{q}}^\infty$.
Here $\eps_{\vec{q}}^\infty$ denotes the energy of the electron far away from the surface.
 
\begin{figure}[t]
\begin{center}
\begin{minipage}{0.5\linewidth}
\includegraphics[width=0.9\linewidth]{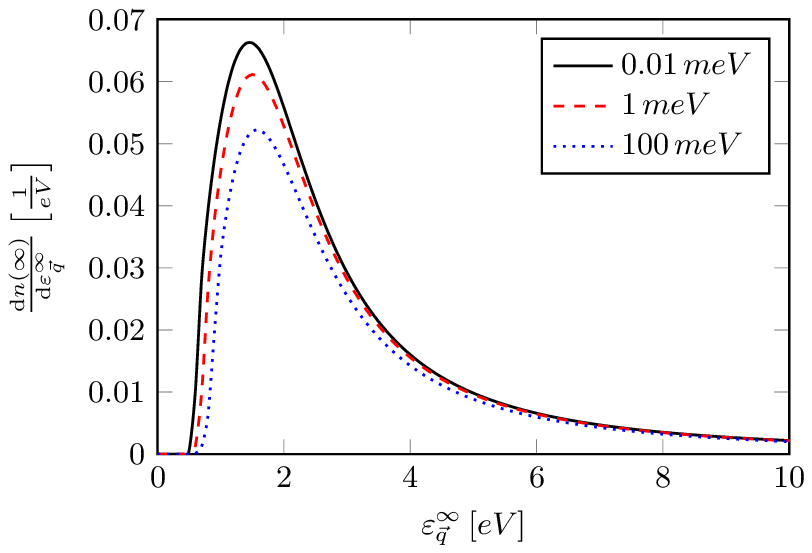}~a)
\end{minipage}\begin{minipage}{0.5\linewidth}
\includegraphics[width=0.9\linewidth]{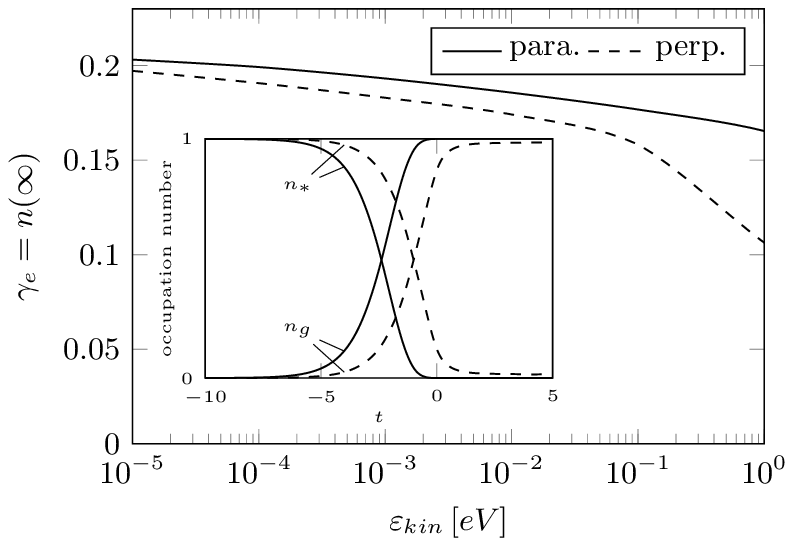}~b)
\end{minipage}
\caption{\textbf{a}) Spectrum of the electron emitted from a \SiOTwo\ surface upon 
impact of a metastable \NitrogenDominantMetastableState\ molecule with perpendicular
molecular axis and three different kinetic energies.
\textbf{b}) Secondary electron emission coefficient $\gamma_e$ due to de-exciting 
\NitrogenDominantMetastableState\ at an \SiOTwo\ surface as a function of the molecule's
kinetic energy and the two orientations of the molecular axis. The inset shows the 
time evolution ($t<0$ incoming and $t>0$ outgoing branch of the trajectory) of the fractions 
of the metastables $n_*$ and ground state molecules $n_g$.} 
\label{fig:5}
\end{center}
\end{figure}
 
Using the above formalism we investigated for various dielectric surfaces the de-excitation
of~\NitrogenDominantMetastableState\ via reaction~(\ref{eq-rct-reaction})~\cite{MBF12}.
Figure~\ref{fig:5} shows results for a \SiOTwo\ surface. The spectrum of the emitted 
electron is shown in Fig.~\ref{fig:5}\textbf{a}. The kinetic energy of the molecule is~$50\,meV$ 
and only the natural decay was taken into account because it dominates the surface-induced 
decay by one order of magnitude~\cite{MBF12}. The cut-off of the spectrum at low energy 
arises from the trapping of the emitted electron in the image potential when its perpendicular 
kinetic energy is too small. The secondary electron emission coefficient $\gamma_e$ 
is shown in  Fig.~\ref{fig:5}\textbf{b} as 
a function of the molecule's kinetic energy~$\eps_{kin}$. The inset, finally, depicts 
for a molecular kinetic energy of~$50\,meV$ the time evolution of the fraction 
of metastables~$n_*$ and ground state molecules~$n_g$. These data demonstrate clearly that 
due to the fast decay of \NitrogenNegativeIonResonance\ the decrease of $n_*(t)$ leads to an 
almost instantaneous increase of $n_g(t)$ by the same amount and thus to a very efficient 
release of an electron. 

It is interesting to note that the de-excitation of~\NitrogenDominantMetastableState\ extracts
an electron from the valence band whereas the wall charge would reside in the conduction band of 
\SiOTwo\ (see Fig.~\ref{fig:4}\textbf{b} for an illustration). The charge transfer due to 
the build-up of the wall charge, on the one hand, and the de-excitation of~\NitrogenDominantMetastableState , 
on the other, thus leads to an inversion of the band occupancies. The time and length scales 
on which it persists until it is eliminated by radiative processes or electron relaxation and transport 
are not yet explored. Support for this microscopic picture comes however from measurements of the wall charge 
in dielectric barrier discharges which indicate the presence of positive wall charges~\cite{Bogaczyk12}. From 
our perspective, these are holes in the dielectric's valence band. 


\section{Concluding remarks}

In a still on-going effort we question the perfect absorber model for plasma walls and 
develop concepts and tools for a microscopic description of the interaction of electrons
and other species with plasma walls. We are particularly interested in how electrons are 
deposited in and extracted from floating plasma walls and how they are distributed across 
the plasma-wall interface once a quasi-stationary floating potential is established. 

So far, we mostly considered uncharged metallic and dielectric surfaces but our 
results already indicate that floating dielectric plasma walls with negative 
electron affinity cannot be described as perfect absorbers because $s_e\ll 1$, irrespective
of the charge of the surface. Our results also indicate that in this case the wall 
charge forms a quasi two-dimensional electron film in front of the surface. The total
charge collected by a plasma wall most probably does not depend on whether it is trapped 
in front or inside the wall; but knowing where the charge resides may be useful for 
developing diagnostics of the wall charge. The time scale, in contrast, on which 
the quasi-stationary wall charge develops should depend on the way it is trapped. A time-resolved
study of, for instance, charging and de-charging of grains in a plasma (or an electro-static 
trap) could thus reveal further insights into the microphysics of wall charges.
As yet unexplored is the electronic inversion temporarily existing
in a floating dielectric wall because the electron influx deposits electrons 
in the conduction band (or in unoccupied surface states, depending on the electron 
affinity), and the influx of ions and/or metastables extracts electrons from the 
valence band. We focused on charge extraction due to metastable nitrogen but the 
inversion also arises due to wall recombination of positive ions. It is thus 
generic for a floating dielectric wall and should be investigated in detail, 
particularly in connection with discharges gliding on floating dielectric 
surfaces~\cite{Bogaczyk12} where the build-up and decay of the inversion could affect,
for instance, the spatio-temporal evolution of the discharge.

Fundamental to any interface is charge transfer. This mantra also holds for plasma 
walls although the rich microphysics associated with it reveals itself only after 
a judicious design of the wall and the discharge. The progress in manufacturing solid-state based
microdischarges is therefore particularly encouraging. We used the electron surface
layer to describe a floating plasma wall. Supplemented by an external bias, it can
be however also used for a microscopic description of biased walls, for instance, 
the plasma bipolar junction transistor~\cite{TWH11}, where the plasma wall is an 
integral part of the device. 


\begin{acknowledgement}
Support from the Deutsche Forschungsgemeinschaft through the TRR 24 
is greatly acknowledged. J. M. was funded by the federal state of 
Mecklenburg-Western Pomerania through a postgraduate scholarship.
\end{acknowledgement}

\bibliographystyle{cpp}
\bibliography{Interface}

\end{document}